\documentclass[apjl]{emulateapj}

\usepackage{natbib}
\usepackage{mathtools}
\usepackage{amssymb}
\usepackage{hyperref}
\usepackage{cleveref}
\crefname{figure}{Fig.}{Figs.}
\newcommand{\refereebf}[1]{{#1}}

\newcommand{\Msun}{ \ensuremath{M_{\odot}} }
\newcommand{\Mpc}{ \ensuremath{{\rm Mpc}} }

\newcommand{\equ}[1]{\cref{eq:#1}} 

\newcommand{\fig}[1]{\cref{fig:#1}}
\shorttitle{Virial shocks in Filaments}
\shortauthors{Birnboim, Padnos \& Zinger}

\begin{document}

\newpage

\title{The hydrodynamic stability of gaseous cosmic filaments}

\author{Yuval Birnboim\altaffilmark{1}, Dan Padnos and Elad Zinger}
\affil{Racah Institute of Physics, The Hebrew University, Jerusalem 91904,  Israel}
\altaffiltext{1}{Research School of Astronomy \& Astrophysics, Australian National University, Canberra, ACT, Australia}
\email{yuval@phys.huji.ac.il}

\label{firstpage}
\begin{abstract}
Virial shocks at edges of cosmic-web structures are a clear
prediction of standard structure formation theories. We derive a
criterion for the stability of the post-shock gas and of the virial
shock itself in spherical, filamentary and planar infall
geometries. When gas cooling is important, we find that shocks become
unstable, and gas flows uninterrupted towards the center of the
respective halo, filament or sheet. For filaments, we impose this
criterion on self-similar infall solutions. We find that instability
is expected for filament masses between
$10^{11}-10^{13}\Msun\,\Mpc^{-1}.$ Using a simplified toy model, we
then show that these filaments will likely feed halos with
$10^{10}M_{\odot}\lesssim M_{halo}\lesssim 10^{13}M_{\odot}$ at
redshift $z=3$, as well as $10^{12}M_{\odot}\lesssim M_{halo}\lesssim
10^{15}M_{\odot}$ at $z=0.$

The instability will affect the survivability of the filaments as they
penetrate gaseous halos in a non-trivial way. Additionally, smaller
halos accreting onto non-stable filaments will not be subject to
ram-pressure inside the filaments.  The instreaming gas will continue
towards the center, and stop either once its angular momentum balances
the gravitational attraction, or when its density becomes so high that
it becomes self-shielded to radiation.

\end{abstract}

\section{Introduction} 
The thermodynamic state of gas in cosmic web filaments has important
implications for observations and theoretical predictions. It will
affect halos which are fed by those filaments, as well as small halos
that accrete onto those filaments as part of the cosmic hierarchical
growth.

Spherical virialization of gas in halos has been a prediction of
galaxy formation models for decades. In particular, it has been shown
\citep{ro77,silk77,binney77,white91} that a comparison between cooling
times and ages of galactic halos can predict the transition from
galaxies to groups and clusters. \citet[hereafter BD03]{bd03} derived
a stability criterion against gravitational collapse of the gas in the
presence of significant cooling. They find that for halos below
$M_{crit}\simeq 10^{12}\Msun$ a hot gaseous halo is not expected to
form, and gas will free-fall until it reaches the disk, at which point
it will stop, radiating its kinetic energy abruptly at that
point. This has been confirmed in multiple hydrodynamical simulations
\citep[e.g.\@ ][]{keres05,ocvirk08,cafg11} and successfully reproduces
star forming galaxies at high-z \citep{db06,dekel09} and the
color-magnitude bi-modality
\citep{db08,croton06,cattaneo06}. Observational
indications of this scenario are
gradually accumulating \citep[e.g.\@][]{dijkstra09,kimm10,martin15}.

In this letter we derive a criterion for the stability of virial shocks around filaments and sheets that form the cosmic web, 
analogous to the BD03 criterion for halos. Following \citet[hereafter
  FG84]{fg84}, we construct self-similar density profiles of filaments.  We apply our stability analysis
to these profiles to identify filaments around which a stable virial shock is
expected to form. This criterion is translated to a more useful form
by identifying which halos are expected to be fed by these
filaments. We find that filament instability influences a
large portion of halos in the universe throughout cosmic age.

The stability of filaments has been addressed before, numerically
\citep{harford11} and analytically \citep{freundlich14,breysse14}, but
without taking into account cooling, and by analyzing the stability of
initially static filaments, ignoring the effects of the shock at the
filaments' edge.

In \S~2 we derive the stability criterion for the existence of
virialized gas in 1,2 and 3 dimensional collapse. In \S~3 we
relate our local criterion to cosmic filaments according to
the self-similar solutions of FG84. In \S~4 we relate
these filaments to typical halo masses that will likely be fed by
them. In \S~5 we summarize and conclude.
 
\section{virial shock stability in spherical, cylindrical and planar geometry}
\label {sec:analytics}
The analysis in BD03 was performed for spherical accretion in the presence of cooling. In this section we generalize that derivation for
infall onto spherical,
cylindrical (onto filaments), and planar  (onto sheets or
disks) objects.

The ideal gas equation of state (EoS) is
\begin{equation}
P=(\gamma-1)\rho e,
\label{eq:ideal}
\end{equation}
with
$\gamma=\left(\frac{\partial~ {\rm ln}P}{\partial~
    \rm{ln}\rho}\right)_s$ the adiabatic index, and $\rho,e,P$ the density, internal energy
and pressure respectively. 
The adiabatic index measures the ``stiffness'' of the EoS, or the adiabatic pressure response to compression. By analogy, we define the effective EoS index, $\gamma_{\rm eff},$ of a parcel
of gas undergoing compression as its pressure response  along a Lagrangian trajectory:
\begin{equation}
\gamma_{\rm eff} \equiv
\frac{d~\rm{ln}P}{d~\rm{ln}\rho}=\frac{\rho}{P}\frac{\dot{P}}{\dot{\rho}},
\label{eq:geff}
\end{equation}
with upper dot implying a Lagrangian time derivative.
\refereebf{Following BD03, this form is transformed into}
\begin{equation}
\gamma_{\rm eff}=\gamma-\frac{\rho}{\dot{\rho}}\bigg{\slash} \frac{e}{q}.
\label{eq:geff2}
\end{equation}
\refereebf{with $q$ related to the \citet{sutherland93} cooling rate according to}
\begin{equation}
q=\rho\Lambda_{\rm cool}(T,Z)=\rho(N_a^2
\chi^2/\mu^2)\Lambda_{\rm mic}(T),
\label{eq:lambda}
\end{equation}
\refereebf{
where ${N_a}/{\mu}$ is the
  average number of molecules per unit mass and $\chi$ is the number
  of electrons per particle. This parametrization allows for altered ionization state of the gas \citep[see ][]{cantalupo10}.
 This equation illustrates how the effective stiffness of the gas is set by two competing timescales: the compression time, $\frac{\rho}{\dot{\rho}}$, and the cooling time, $\frac{e}{q}$. If radiative losses are significant during a compression time, we get effective softening, $\gamma_{\rm eff}<\gamma$. In the absence of cooling, we recover $\gamma_{\rm eff}=\gamma$.}

The compression time for post-shock infalling gas depends on the velocity and geometry of the infall through the Lagrangian continuity equation,
\begin{equation}
\dot{\rho}=-\rho \nabla \cdot u=-\rho\left((n-1)\frac{u}{r}+\frac{\partial
  u}{\partial r}\right),
\label{eq:nabla}
\end{equation}
with $u$ being the velocity of the gas, and $n$ the dimensionality of
the infall: $n=1,2,3$ corresponds to planar, filamentary and spherical collapse respectively. Assuming that the post-shock flow is
homologous,
\begin{equation}
u=\frac{u_s}{r_s}r
\label{eq:homo}
\end{equation}
with $r_s$ the shock radius and $u_s$ the velocity directly below the shock, we can apply \equ{nabla} to show  that the post-shock gas must contract uniformly, independent of $r$,
\begin{equation}
\frac{\dot{\rho}}{\rho}=-n\frac{u_s}{r_s}={\rm Const}.
\label{eq:rhodot}
\end{equation}
This homologous behavior is seen in 1D simulations for the spherical
case \citep{bd03}, and is a standard assumption in hydrostatic
stability calculations. It also roughly matches self-similar
solutions of gaseous spherical infall \citep{bertschinger85}. 

We consider a quasi-static configuration where the shock radius, $r_s$, and the post-shock velocity profile \equ{homo} are approximately constant. We test the stability of this configuration by assuming that it is initially hydrostatic,
\begin{equation}
\ddot{r}=-\frac{1}{\rho}\nabla P+a_g=0
\label{eq:rddot}
\end{equation}
($a_g$ being the gravitational acceleration)
but has an inward post-shock velocity, and checking the sign of
the auxiliary
acceleration or ``jerk'', $\dddot r$, that forms due to this motion.
A positive jerk
will instigate outwards motion implying a stable configuration,
while a negative $\dddot{r}$ will lead to collapse, implying instability.\footnote{The hydrostatic condition, \equ{rddot}, assumes the acceleration implied by the homologous velocity profile can be considered negligible compared to the deceleration across the shock. This is a good approximation for strong shocks (see \S~3). We avoid the inclusion of a homologous acceleration term because it complicates the derivation considerably but yields only an insignificant quantitative correction.}

The gravitational acceleration depends on the dimensionality of
the potential well. Since the perturbation is Lagrangian, the
mass enclosed below the gas parcel is constant, so
\begin{equation}
a_g=-Ar^{1-n},
\label{eq:ag}
\end{equation}
with $A$ a positive constant\footnote{for spherical, cylindrical and
  planar configuration the gravitational acceleration is
  $-GM/r^2,-2Gl/r,-2\pi G \Sigma$ respectively, with $G,M,l,\Sigma$ constants.}.
We convert the $\nabla=\frac{d}{dr}$ operator in \equ{rddot} to a Lagrangian
mass derivative, by noting that a mass element is related to a spatial
differential according to:
\begin{equation}
dm=B\rho r^{n-1}dr,
\label{eq:dm}
\end{equation} 
with $B$ a positive constant\footnote{Note that the units of $dm$ here
  depend on the dimensionality of the infall, $n$: it is mass for
  spherical infall, mass per unit length for cylindrical infall
  and mass per unit area for planar infall.}.
Plugging \cref{eq:ag,eq:dm} into \equ{rddot} we find
\begin{equation}
\ddot{r}=-Br^{n-1}\frac{dP}{dm}-Ar^{1-n}=0.
\label{eq:rddot2}
\end{equation}
From here on we shall denote all mass derivatives with $'$.

The rate of change in the acceleration is the time derivative
of \equ{rddot2},
\begin{equation}
\dddot{r}=-B(n-1)r^{n-2}uP'-Br^{n-1}\dot{P'}-A(1-n)r^{-n}u,
\end{equation}
noting that $u=\dot{r}.$
Eliminating the last term by use of \equ{rddot2}, collecting terms,
and exchanging the spatial and time derivative of $P$
one gets: 
\begin{equation}
\dddot{r}=-Br^{n-1}[2(n-1)\frac{u}{r}P'+(\dot{P})'].
\label{eq:dddr}
\end{equation}

The calculation of $(\dot{P})'$ is somewhat lengthy. We first derive
an expression for $\dot{P}$:
\begin{align}
  \dot{P}&=\gamma_{\rm eff}\frac{\dot{\rho}}{\rho}P=\gamma\frac{\dot{\rho}}{\rho}P-\frac{q}{e}P
  \nonumber\\
  &=\gamma\frac{\dot{\rho}}{\rho}P-\frac{\rho P
  \Lambda_{\rm cool}}{e}=\gamma\frac{\dot{\rho}}{\rho}P-(\gamma-1)\rho^2\Lambda_{\rm cool},
\label{eq:pdot2}
\end{align}
with the first equality due to \equ{geff}, the second equality to
\equ{geff2}, third to \equ{lambda} and fourth to \equ{ideal}.
By \equ{rhodot} we note that the term $\frac{\dot{\rho}}{\rho}$ is
independent of the spatial derivative. We further assume that the
cooling function, $\Lambda_{\rm cool},$ does not change significantly due to the
change of temperature of the Lagrangian mass element \footnote{This assumption is reasonable for $\Lambda_{\rm cool}(T,Z),$ except near $10^4K,$ and is necessary for an analytic solution to be possible. We neglect it here, at the risk of a slight error near the lower boundary of the unstable regime.}, so $\Lambda_{\rm cool}$ can
also be taken out of the derivative. We assume that nearby mass elements directly below the shock start
with the same thermodynamic conditions (i.e.\@ they lie on the same adiabat). Combined with the definition in \equ{geff}, this indicates that
\begin{equation}
\frac{P'}{\rho'}=\gamma_{\rm eff}\frac{P}{\rho}.
\label{eq:ptag}
\end{equation}
Differentiating \equ{pdot2} then yields:
\begin{align}
&(\dot{P})'= \label{eq:pdottag}\\
&\gamma \frac{\dot{\rho}}{\rho}P'-2(\gamma-1)\rho
\Lambda_{\rm cool}\rho'=\gamma
\frac{\dot{\rho}}{\rho}P'-2(\gamma-1)q\frac{1}{\gamma_{\rm eff}}\frac{\rho}{P}P'=\nonumber\\
&\gamma
  \frac{\dot{\rho}}{\rho}P'-\frac{2}{\gamma_{\rm eff}}\:\frac{q}{e}P'=\gamma
  \frac{\dot{\rho}}{\rho}P'-\frac{2}{\gamma_{\rm eff}}(\gamma-\gamma_{\rm eff})\frac{\dot{\rho}}{\rho}P'.\nonumber
\end{align}
The second equality is due to \equ{ptag}, the third by reverse use of
\cref{eq:ideal,eq:lambda}, and the fourth by separating
$\frac{q}{e}$ from \equ{geff2}.

Inserting \equ{pdottag} into \equ{dddr}, and converting
$\frac{\dot{\rho}}{\rho}$ and $\frac{u}{r}$ to $\frac{u_s}{r_s}$
according to \cref{eq:homo,eq:rhodot}, we finally get:
\begin{equation}
\dddot{r}=-Br^{n-1}\frac{u_s}{r_s}P'[2(n-1)-n\gamma+\frac{2n}{\gamma_{\rm eff}}(\gamma-\gamma_{\rm eff})].
\end{equation}
$u_s<0$ because below the standing shock there is inwards velocity, and
$P'<0$ so that the pressure force, $-\nabla P,$ is positive to balance gravitation in the quasi-hydrostatic halo, so the factor before the square brackets is always negative. A
positive jerk, or stability, thus occurs when
\begin{equation}
2(n-1)-n\gamma+\frac{2n}{\gamma_{\rm eff}}(\gamma-\gamma_{\rm eff})<0.
\label{eq:stability}
\end{equation}
We first note that in the absence of cooling $\gamma_{\rm eff}=\gamma$ and
the stability condition reduces to:
\begin{equation}
\gamma>2-\frac{2}{n}.
\end{equation}
For the spherical case ($n=3$) the stability criterion is
$\gamma>\frac{4}{3},$ recovering a well known result. For filamentary
accretion ($n=2$) of adiabatically collapsing gas stability is gained when
$\gamma>1$, and for planar collapse $(n=1)$ when $\gamma>0.$

When cooling is present the stability criterion is:
\begin{equation}
\gamma_{\rm eff}>\frac{2n\gamma}{n\gamma+2},
\end{equation}
which, for monoatomic gas ($\gamma=\frac{5}{3}$) is
$\frac{10}{7},~\frac{5}{4},~\frac{10}{11}$ for spherical, cylindrical
and planar collapse respectively.

We note that for $\gamma=\frac{5}{3}$ the critical value of
$\gamma_{\rm eff}$ in the presence of cooling is always somewhat larger
than the adiabatic critical value. Hence, monoatomic gas that is
cooling with some local $\gamma_{\rm eff},$ (with entropy and energy decreasing due to cooling) is always less stable
than an adiabatic gas with a softened EoS
$\gamma=\gamma_{\rm eff}$.

\section{Stability of cosmological filaments}
\label {sec:selfsimilar}

\begin{figure}
\begin{center}
\includegraphics[width=3.3in]{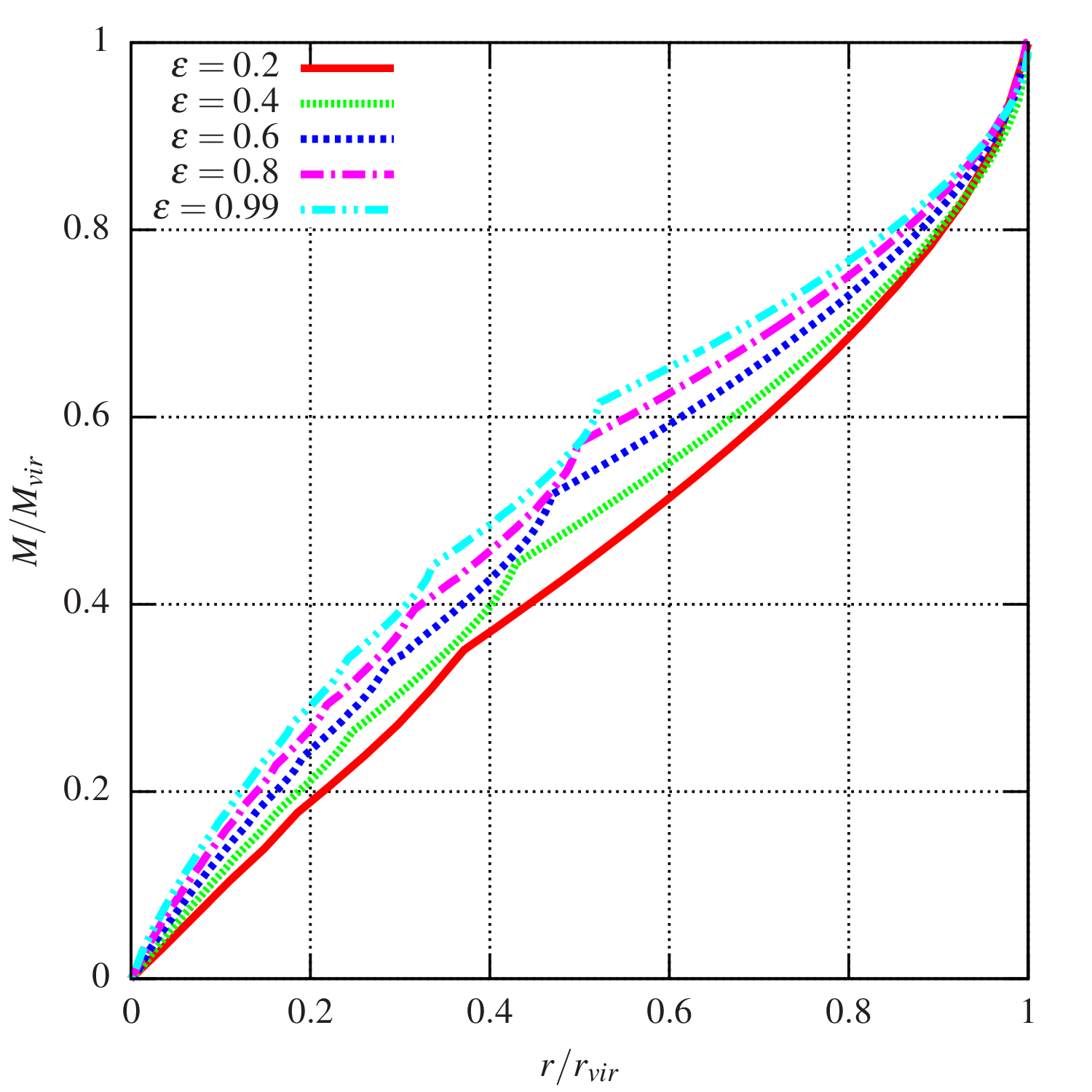}
\caption{\label{fig:massnorm} Mass profiles of self-similar
  cylindrical accretion normalized to the virial radius and virial
  mass, as defined by the first shell crossing, for various initial
  perturbation power-laws coefficients $\epsilon$.}
\end{center}
\end{figure}

\begin{figure}
\begin{center}
\includegraphics[width=3.3in]{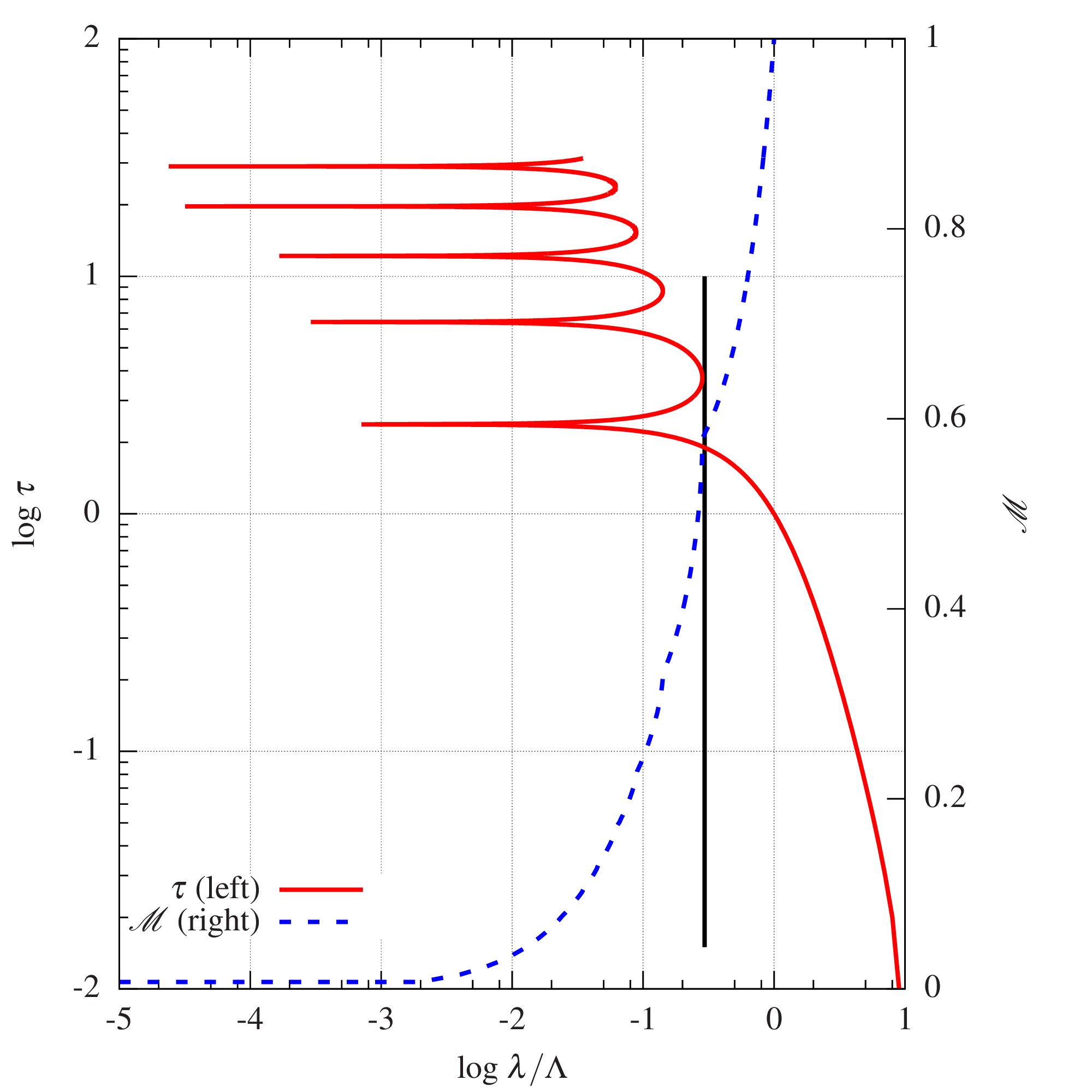}
\caption{\label{fig:ml} Self-similar DM
  trajectory and mass profile of an $\epsilon=0.8$ perturbation. At
  time $t_*$ a specific shell with mass $M_*$ and radius $x_*$ is at
  turnaround. The unitless time is $\tau=t/t_*$ and unitless radius is
  $\lambda=x/x_*$\refereebf{ (for consistency, all variables are named according to FG84)}. {\it Red (left axis):} the self-similar shell
  trajectory with $\lambda$ normalized to the current turnaround
  radius at each time $\Lambda(\tau).$ {\it Blue (right axis):} the
  self-similar mass profile in units of $M_*.$ The black vertical line
  corresponds to the radius at which shell crossing first occurs, and
  is evident in the mass profile as well as in the trajectory.
}
\end{center}
\end{figure}

Filaments grow (in girth) by accreting gas from their
surroundings. The infall geometry is primarily cylindrical, although
most of the gas is channeled along sheets, and becomes spherical in
the vicinity of halos. The flow parallel to the filament below and
above the shock is continuous, and can be factored out locally with a
proper shift of the frame of reference. Dark matter (DM) and gas
accrete onto filaments together. Separation between gas and DM occurs
when gas becomes thermalized and is decelerated due to its pressure.
We wish to determine where and how this gas is thermalized,
particularly in the presence of cooling, which softens the effective
EoS of the gas. To do so, we must connect global properties of
filaments: their mass per unit length and their density and velocity
profiles, to local cooling and contraction rates that determine
the stability of the gas.

An ideal framework for connecting the large-scale properties of
filaments to local conditions is the self-similar solutions of
FG84.  We do not present here the derivation and results, and refer
the reader to the original paper. Following FG84, we numerically solve
for the self-consistent density profile and trajectories of infalling cylindrical
DM shells, starting from an initial mass
perturbation within an Einstein de-Sitter universe. In this framework,
a filament is characterized by its mass per unit length, and by its
initial perturbation: 
\begin{align}
M_{fil}(r)&=M_{fil}^0(r)\left(1+\frac{\delta M_{fil}(r)}{M_{fil}^0(r)}\right)\nonumber\\
M_{fil}^0(r)&\equiv\pi r^2\rho_u,
\end{align}
with $\rho_u$ the universal density at the initial time. The perturbation is defined as a function of the unperturbed mass:
\begin{equation} 
\frac{\delta M_{fil}}{M_{fil}^0}=\left(\frac{M_{fil}^0}{M_*}\right)^{-\epsilon},
\end{equation}
with $M_*$ a reference mass.
 $\epsilon$ varies between $0$ and $1$, where $1$ corresponds to the
most localized perturbation that still grows with $M_{fil}^0$, and $0$
to a long range perturbation for which the density of the perturbation
is still decreasing with $M_{fil}^0.$ \fig{massnorm} shows the
resulting mass profiles (normalized to the virial radius and the
virial mass) for various values of $\epsilon.$ It is evident that the
profile within the virial radius of the filament depend weakly on
$\epsilon.$ The visible discontinuities in gradient correspond to
caustics in DM shells as they turn around consecutively. The outermost
caustic, or ``first shell crossing'' is defined as the virial
radius. The mass profiles in \fig{massnorm} are normalized to that
radius.

\fig{ml} shows the self-similar trajectory of a DM shell and the
filament's mass profile. The trajectory's radius is normalized to the
turnaround radius at each time, so before turnaround, at $\tau<1,$ the
spatial coordinate $\lambda/\Lambda>1.$ The mass profile as a function
of that same spatial variable is also present. The vertical black line
corresponds to the event of first crossing (the virial
radius), and is roughly where the virial shock will occur for a
gaseous shell. 

\begin{figure}
\begin{center}
\includegraphics[width=3.3in]{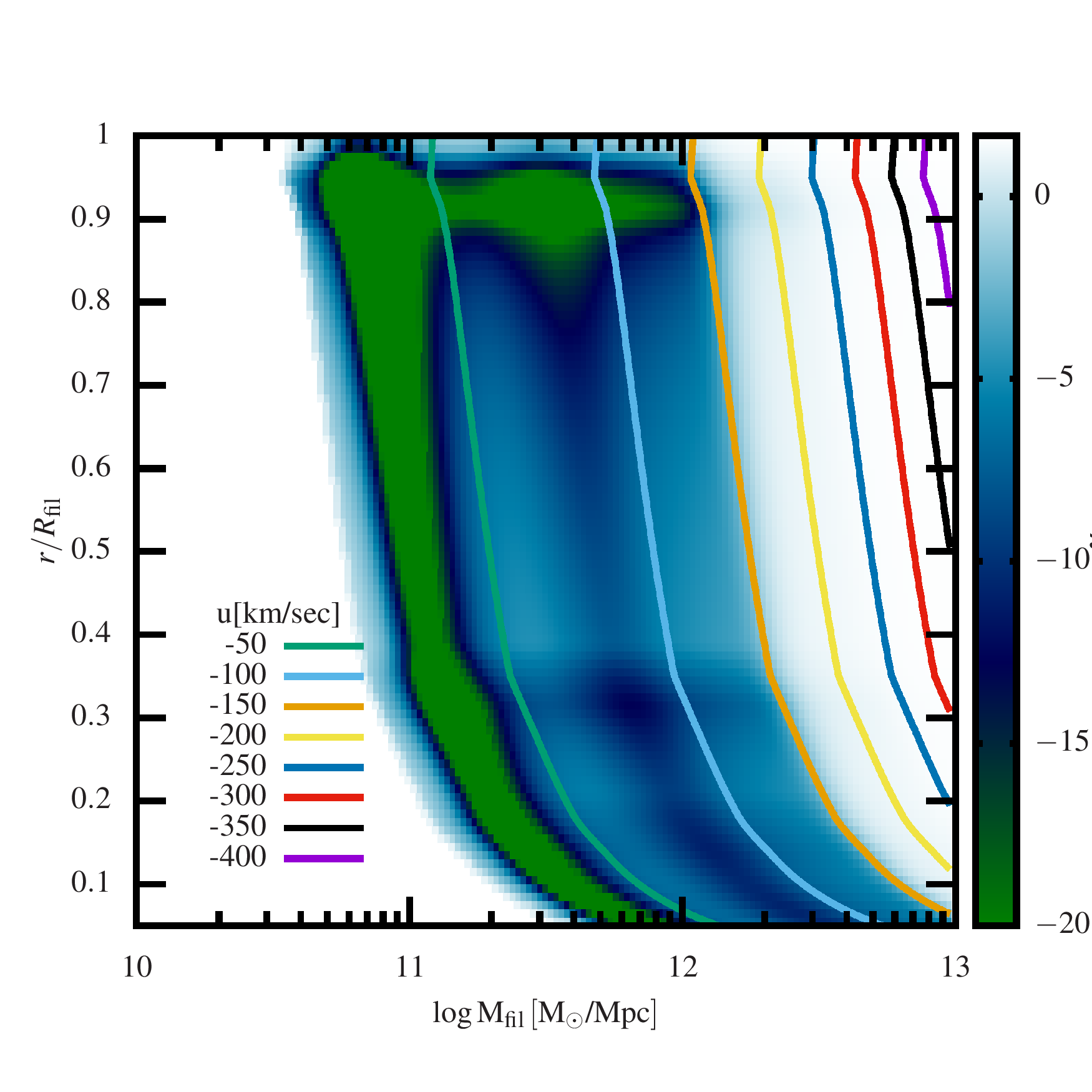}
\caption{\label{fig:gmx_0.2} Stability of $\epsilon=0.2$ perturbation
  as a function of filament mass and radius with respect to the
  filamentary virial radius. {\it colormap:} $\gamma_{\rm eff}$ - white color corresponds to $\gamma_{eff}=\gamma=5/3.$ All
  values below $1.25$ are unstable. {\it contours:} the infall
  velocity.}
\end{center}
\end{figure}

\begin{figure}
\begin{center}
\includegraphics[width=3.3in]{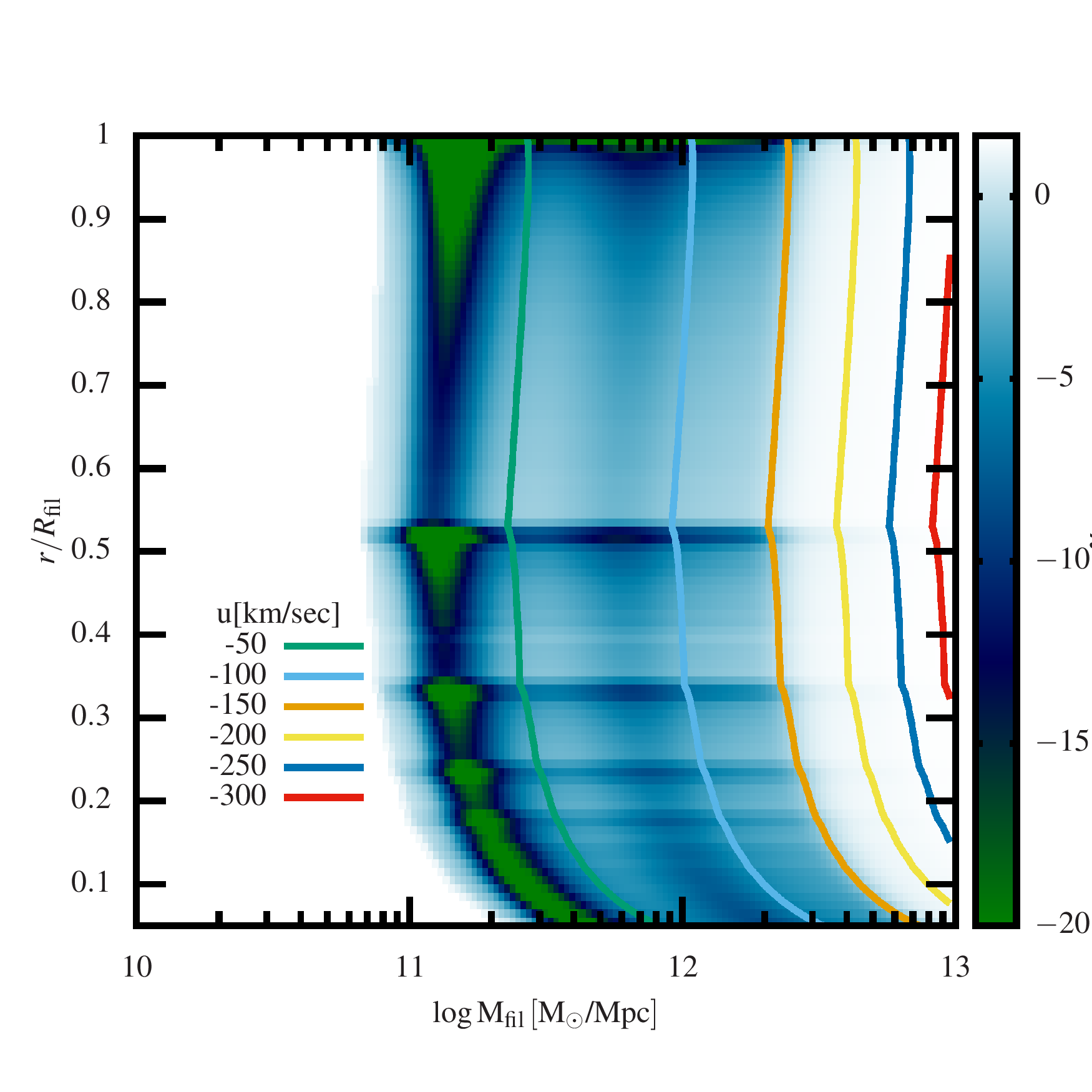}
\caption{\label{fig:gmx_0.99} Same as \cref{fig:gmx_0.2}, but for $\epsilon=0.99$}
\end{center}
\end{figure}

From the self-similar solution we extract the infall velocity and
density at every radius. The infalling velocity and density are for
DM trajectories and correspond to infalling gaseous shells only before
they pass through the virial shock, at which point their pressure
becomes significant and their trajectories diverge from those of the
DM. The stability of the post-shock gas depends on the local compression
rate, density and temperature for the {\it post-}shocked gas. These
values are approximated from the {\it pre}-shocked ones by use of the
strong shock approximation, that is valid as long as the pre-shocked
velocity is much larger than the pre-shocked speed of sound
($c_{s}^0$). Assuming that the $c_s^0\lesssim 10 \mathrm{km\,s^{-1}},$ we show later
  (\cref{fig:gmx_0.2,fig:gmx_0.99}), that this
  approximation goes from being marginally satisfied for
  the smallest filaments to being fully justified for the large
filaments. Using the full shock conditions will not change the results
significantly, and requires knowledge of the thermodynamic state of
the cold gas. As the effective EoS becomes softer, the virial shock
ceases to expand, and starts to collapse. At its
critical state, we expect the shock to be at rest. Using the strong
shock approximation and assuming the shock is at rest, the post-shock
values are:
\begin{align}
&\rho_1=\frac{\gamma+1}{\gamma-1}\rho_0\\
&u_1=\frac{\gamma-1}{\gamma+1}u_0\nonumber\\
&e_1=\frac{2u_0^2}{(\gamma+1)^2}\nonumber\\
&T_1=\frac{\mu(\gamma-1)}{N_ak_B}e_1,\nonumber
\end{align}
with subscript $0$ and $1$ denoting pre-shocked and post-shocked
variables respectively, $\rho_0,u_0$ given from the numerical solution
of the self-similar collapse, and $e_1,T_1$ are the internal energy
and temperature of the post-shock variables. $\mu=0.61$ is the mean
molecular weight for primordial, fully ionized gas, and $N_a,k_B$ are
Avogadro's number and Boltzmann's constant. The
conversion from DM density to gas density is achieved by multiplying
the density by a universal baryonic fraction ($f_b=0.17$ throughout
this work). This value is reasonable as long as the DM and gas flow
together, i.e.\@ for pre-shocked gas.

For a filament characterized by $M$ and $\epsilon$ we calculate
the post-shock values for every radius below $r_{\rm vir},$ and use
\equ{rhodot} and \equ{lambda} inserted into \equ{geff2} to calculate
$\gamma_{\rm eff}$.  \cref{fig:gmx_0.2,fig:gmx_0.99}
show $\gamma_{\rm eff}$ as a function of $M_{fil}$ and $r/R_{\rm
  vir},$ with $R_{\rm vir}$ defined as the radius of first shell crossing of the self-similar solutions (see \cref{fig:ml}). In regions where $\gamma_{\rm eff}$ drops below the threshold
for stability, $\gamma_{\rm eff}<\gamma_{crit}=1.25$, the filament
cannot sustain a virial shock. Moreover, if a shock were to form in regions
where $\gamma_{\rm eff}<0,$ the post-shock pressure would decrease
even as it contracts. The mass range for which the filament
is unstable at least at some radius for both values of $\epsilon$ is
between $10^{11}-10^{13}\Msun\,\Mpc^{-1}.$ In these plots, radii of density caustics leave horizontal features, and peaks in the cooling curve create features parallel to infall velocities. For each radius, the sharp drop in $\gamma_{eff}$ as mass exceeds the lower threshold for instability is due to the post-shock temperature exceeding $10^4K,$ where the cooling rate grows by many orders of magnitude.   

\section{Implication for accretion onto halos}
\label {sec:halos}
\begin{figure}
\begin{center}
\includegraphics[width=3.3in]{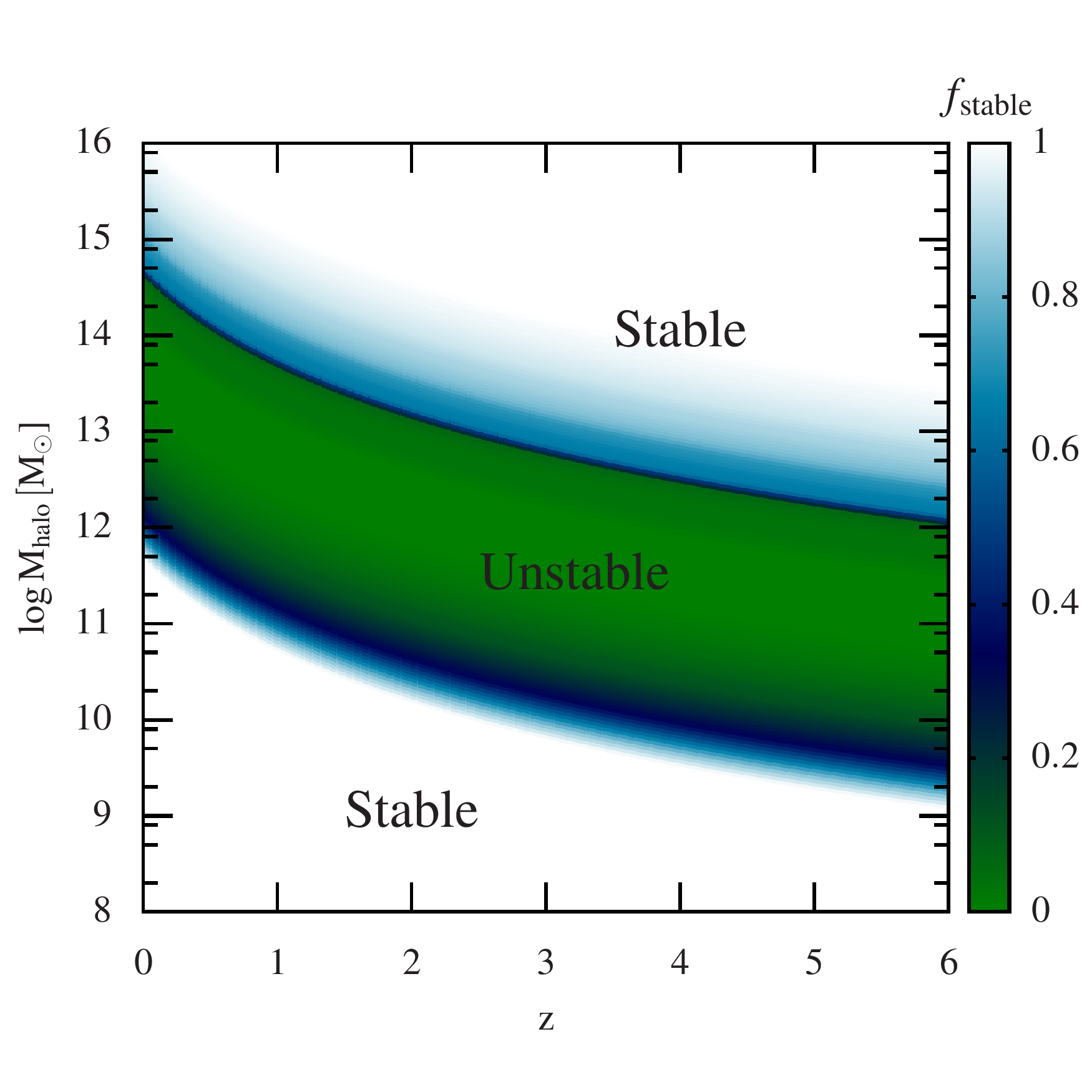}
\caption{\label{fig:stab_0.2} Stability of filaments falling into
  halos as a function of halo mass and redshift, for $\epsilon=0.2.$ {\it colormap:} the stable filament fraction (see text).}
\end{center}
\end{figure}
We now wish to relate the filament masses (per unit length) to the typical masses of halos fed by such filaments. A
full analysis of the filament distribution that accrete onto certain
halos requires cosmological N-body simulations. For simplicity,
we choose an alternative avenue that approximates the relation between
halos and their filaments. \citet{danovich12} study the filamentary
nature of mass accretion onto high redshift galaxies. They find that
typically, halos that originate from high-$\sigma$ peaks in the
initial perturbation accrete most ($f_{acc}\simeq70\%$) of their mass
in filaments, out of which, $95\%$ originates from the combined flow
in the $3$ largest filaments. Although the work analyzes high redshift
galaxies, we expect the same to be true for low-redshift clusters, who are also high-$\sigma$ peaks. Using these values, we estimate the
typical accretion rate through each filament as
\begin{equation}
\dot{M}_{acc}=\frac{f_{acc}}{3}\dot{M}_{halo},
\end{equation}
with $\dot{M}_{acc},\dot{M}_{halo}$ the mass flow rate of gas within a
filament and halo accretion rate
respectively. For the halo accretion rate we use the fit from
\citet{neistein06},
\begin{equation}
\frac{\dot{M}_{halo}}{M_{halo}}= \left(\frac{0.03}{Gyr} \right) \left(1+z\right)^{2.5},
\end{equation}
with $z$ the redshift. Finally, the mass of the filament per unit
length is related to the flow rate by
$M_{fil}=\dot{M}_{acc}/v_{\rm vir}(M_{halo},z)$, using the halo virial
velocity, $v_{\rm vir},$ as an estimate for filaments' velocity as they
accrete onto halos. A typical filament feeding a high-$\sigma$ peak halo of mass
$M_{halo}$ will thus have an estimated mass of:
\begin{equation}
M_{fil}\simeq \frac{f_{acc}}{3} \left(\frac{0.03}{Gyr} \right) \left(1+z\right)^{2.5}\frac{M_{halo}}{v_{\rm vir}}.
\end{equation}

In \fig{stab_0.2} we use the inverse of this transformation to show the stable fraction of a filament, $f_{stable},$ as a function of the halo mass and redshift:
\begin{equation}
f_{stable}\equiv \frac{1}{R_{\rm vir}}\int_0^{R_{\rm vir}}\Theta\left[\gamma_{\rm eff}(r)-1.25\right]dr,
\end{equation}
with $R_{\rm vir}$ the virial radius of the filament, and $\Theta$ the Heaviside function.
From \cref{fig:stab_0.2} it is evident that for $10^{10}M_{\odot}\lesssim M_{halo}\lesssim 10^{13}M_{\odot}$ at $z=3$, as well as for $10^{12}M_{\odot}\lesssim M_{halo}\lesssim 10^{15}M_{\odot}$ at $z=0$ halos are expected to be fed by filaments that are not in hydrostatic stability.

\section{Summary and Discussion}
\label {sec:summary}
We have shown that in the presence of significant cooling, the
accretion process of gas onto cosmic-web structures will not always
proceed according to the standard virialization scenario of the
infall-heating-cooling sequence. 
The analysis shown
here can be applied for accretion onto spherical halos, cylindrical filaments and planar sheets. 
For filament of $10^{11}-10^{13}\Msun\,\Mpc^{-1}$, we show that
gas is expected to fall without ever passing a shock, resulting in
dense, thin filaments with low entropy. This is in complete analogy
to spherical cold accretion onto halos that have been shown in BD03
and demonstrated in observations and simulations. 

Using a simplified toy model for the relation between halo mass and
redshift to typical filaments that feed it, we show that throughout
cosmic history galaxies and clusters are affected by that
instability. In particular, high-z star forming galaxies
($M_{halo}>10^{10}\Msun$ at $z=3$), and low redshift groups and
clusters ($10^{12}\Msun\lesssim M_{halo}\lesssim 10^{15}\Msun$ at
$z=0$) will be fed by filaments for which the gas is unstable.

The process that eventually stops the infall is still
unclear, and we postulate that it is either angular momentum support
from an original helicity of the filament, or by reduction of the
cooling rate due to self-shielding of the gas.  A prediction of this
work is thus that filament gas, in the non-stable regime, will be highly
rotating and angular momentum supported. Both processes are hard to identify
in simulations, and have not been examined so far. In their absence,
gas in simulations will flow towards the center of the filament until
it approaches the numerical gravitational smoothing length, at
which point the force will diminish. This
indicates that the density and entropy of gas in unstable filaments are a numerical artifact and will not converge to the right values. This
problem will be examined in future work.

\refereebf{The lack of virialized gas in filaments is expected to significantly affect the outcome of galaxies falling onto the filament, and of halos fed by the filament. Halos falling onto filaments are expected to looe gas through ram pressure stripping, and to enrich the filament with metals. Both these processes will be suppressed when galaxies fall into filaments with no stable atmosphere. Penetration of cosmic-web gas directly to galaxies affects the ISM state, and the gas available for star formation and AGNs, as well as their feedback efficiencies. \citet{mandelker16} analyze the Kelvin-Helmholtz stability of supersonic filaments. They find that filaments lose stability via bulk modes, that correspond to standing waves reflecting through edges of the filament. These results do not account for the effects of gravitational attraction towards the center of the filament, and to angular momentum support, both expected to stabilize the filament further. These effects will be addressed in future work.

Observationally, the temperature of the filament could affect its detectability through Lyman$-\alpha$ absorption \citep{narayanan10,wakker15} and emission \citep{martin15}. The temperature of the filaments will also affect the soft X-ray background and the total amount of gas in the ``warm phase'' \citep{cen99,dave01}. All these effects are left for analysis in future work.}

\acknowledgments We thank Oliver Hahn for useful discussions. YB and DP have been supported by ISF grant 1059/14.

\label{lastpage}
\end{document}